# Imaging and registration of buried atomic-precision donor devices using scanning capacitance microscopy

E. Bussmann*, M. Rudolph*, G. S. Subramania, S. Misra, S.M. Carr, E. Langlois,

J. Dominguez, T. Pluym, M.P. Lilly, and M.S. Carroll

Sandia National Laboratories, PO Box 5800 Albuquerque NM, 87185

We show that a scanning capacitance microscope (SCM) can image buried delta-doped donor nanostructures fabricated in Si via a recently developed atomic-precision scanning tunneling microscopy (STM) lithography technique. A critical challenge in completing atomic-precision nanoelectronic devices is to accurately align mesoscopic metal contacts to the STM defined nanostructures. Utilizing the SCMs ability to image buried dopant nanostructures, we have developed a technique by which we are able to position metal electrodes on the surface to form contacts to underlying STM fabricated donor nanostructures with a measured accuracy of 300 nm.  Low temperature (T=4K) transport measurements confirm successful placement of the contacts to the donor nanostructures.

* These authors contributed equally to this work

**Introduction**

Recently, a scanning tunneling microscopy (STM) based technique to fabricate atomically-precise nanoelectronic devices via phosphorus delta doping in Si or Ge has been developed culminating in the demonstration of a transistor with just a single-donor channel [1-10]. The atomically-precise STM fabrication technique is of growing interest in fields such as quantum computing [11-14], as well as having potential utility for other areas such as in testing the limits of CMOS transistor scaling [7]. In the STM technique, the device's active regions are defined using STM hydrogen depassivation lithography [15-18]. By exposing the patterned surface to a phosphine ambient, phosphorus donors are adsorbed into regions where hydrogen has been desorbed. The dopants are incorporated using a low temperature anneal that decomposes the phosphine molecule on the surface, leaving a substitutional phosphorus atom in the surface [10]. Once the phosphorus is incorporated, an epitaxial silicon growth step is used to bury the donors [4]. The resulting buried donor nanostructure is 1-6 atomic layers thick and can have donor (electron) density >$10^{14}$ cm$^{-2}$ [18,19].

A central difficulty in completing a nanoelectronic device with the STM technique is making mesoscopic metal contacts to the STM-defined donor nanostructures [5,6], since the doped active regions are nano-sized, a few atomic layers thick, and typically buried under 10-to-100 nm-thick epitaxial Si, rendering them optically invisible. The STM part of the process flow, furthermore, is known to be incompatible with standard (W, Ti-Au) metal alignment marks used in electron beam lithography (EBL), making it difficult to achieve high accuracy alignment of surface contacts with the buried donor nanostructure.

In this work, we show that scanning capacitance microscopy (SCM) [20,21] can image buried donor nanostructures produced via the STM technique. We use SCM to locate a buried donor nanostructure with nanoscale accuracy relative to metal alignment features added to the sample surface after the STM process step is complete (i.e., after the buried donor structure is encapsulated in silicon). Once the coordinates of the buried donor structure are known, we align metal contacts (ohmics) to the buried structure. Successful placement of the contact metal is confirmed by transport measurements on the completed device at low temperature (T=4K). In addition, the SCM provides non-destructive *ex situ* spatial metrology for the donor nanostructures, which to date has not been done for the STM fabrication approach, except under special conditions by *in situ* STM [22]. As an example of the metrology application, we show SCM images providing direct visual confirmation of the accurate step-by-step placement of via holes and contact metal layers over the buried donor layer.

**Methods**

We prepare samples following a recipe modeled on that of Simmons et al. and described in detail in previous publications [1-10]. First, we etch an array of optically visible coarse registration marks (100-300 nm deep) into the Si that will serve to align each step of the device fabrication process with few-μm (coarse) accuracy. Prior to the process, a 15-nm-thick sacrificial thermal oxide is grown on the Si(100) (miscut<0.1°, N~$10^{15}$/cm$^3$ B) wafers to clean the surface and getter contaminants. The oxide also serves as a hard mask for marker etching. Then optical lithography and a dry etch are used to transfer the alignment features into the Si. Subsequently, a 50-nm-thick protective thermal oxide is grown. The wafers are then coated in a micron-thick layer of photoresist and diced. Prior to STM processing, die are cleaned using a three step process. First, the photoresist is stripped via a soak in acetone/IPA for 10 min. Then the samples are cleaned using a piranha etch (3:1 $H_2SO_4$:$H_2O_2$, 90°C, 10 min), then an HF dip (10:1 $H_2O$:HF, room temperature, 1 min), and then a chemical reoxidation (5:1:1 $H_2O$:$H_2O_2$:HCl, 60°C 10 min). Samples are rinsed for 10 min in DI $H_2O$ following each step and blasted dry with $N_2$ at the end of the process.

Donor device fabrication is performed with a homebuilt STM system with a base pressure of 5x$10^{-10}$ Torr. Fig. 1 (a-f) shows the process to fabricate a donor nanostructure via STM. The samples are loaded into the STM after the *ex situ* clean and degassed overnight at ~600°C.

We prepare Si(100)-2x1:H monohydride surfaces using a combination of high temperature flash annealing and exposure to atomic hydrogen [3, 15-17]. First, the sample is rapidly annealed T>1200°C several times for 10 s using DC Joule heating. The pressure in the STM system does not exceed 2x$10^{-9}$ Torr during the anneal. The flash anneal sublimates surface oxide, the Si itself, as well as many species of contaminants. STM measurements after the flash reveal a stepped 2x1 reconstructed surface with <1% density of common surface defects, primarily dimer vacancies and c-type defects. Finally, the sample surfaces are H-terminated by exposure to a background pressure of $H_2$ ($10^{-6}$ Torr) with a hot tungsten filament (~2000°C) placed within a few centimeters of the sample, which is held at 350°C during the process. Samples are then transferred to the STM stage for lithography.

The STM tip is moved to the sample while viewing the tip position with an optical microscope, Fig. 1 (g). In our technique, the device nanostructure can be placed at any arbitrary location of choice on the sample. We typically choose to write devices after positioning the tip within 100 um of some etch feature, as in Fig.(g,h). Numerous etch features are arrayed in 5x5 mm field accessible to the tip, providing ample space for device patterning.

Device patterns are written by STM hydrogen depassivation lithography in a hybrid mode in which atomic-precision features are written at relatively low tip-substrate voltages (5-7V), while larger micron-scale features are written in a high-voltage field-emission regime (V~10-100V) where depassivation is very rapid, but linewidths are 10-100 nm [15-17]. Fig. 1 (i,j) shows the STM lithography pattern of a four-terminal single-electron transistor written in this hybrid mode. The four 1x1.5 µm$^2$ rectangles serve as contact pads to the donor layer, while the active region of the device, shown in the inset, consists of a channel with a single 20-nm-sized quantum dot (I) bridging the source (S) and drain (D) leads. Two other donor regions serve as gates (G1, G2) intended to control electron occupation of the dot.

The electrically active donor nanostructure is formed by doping the depassivated regions by exposing the sample to a background pressure of $PH_3$ ($2\times10^{-8}$ Torr for 5 min) following the depassivation step, Fig. 1 (d). The $PH_3$ adsorbs on the depassivated regions. Phosphorus incorporation into the Si is induced by annealing the sample at 350°C for 10 s, Fig. 1 (e). Phosphorus incorporation is constrained to the depassivated regions by the remaining surrounding hydrogen resist. The donor layer is then capped with epitaxial Si from a sublimation source to bury the donor structure, Fig. 1 (f). The Si epitaxial layer is typically 10 to 100 nm thick. The sample is held at the relatively low temperature of 250°C during the epitaxy step to limit surface segregation of the donors [16,17].

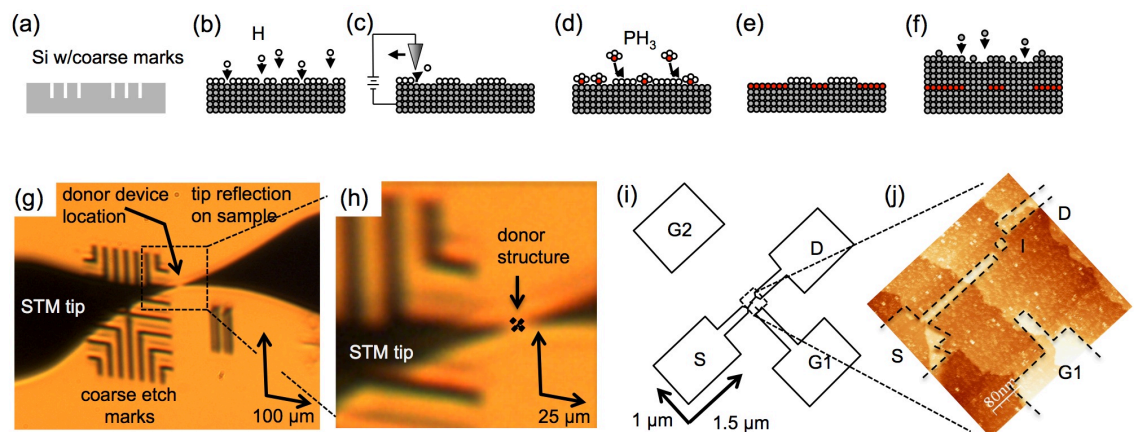

**Figure 1 (a-f) Process for STM fabrication of buried donor nanostructures. (g,h) Optical images of the STM tip and coarse etched alignment marks *in situ* during fabrication. (i) Schematic for a four-terminal single-electron transistor device showing the source (S), drain (D) and two gates (G1,2). (j) STM image of the hydrogen depassivation lithography pattern that forms the active channel of the SET device.**

The buried delta-doped structures have donor and electron densities $N_D$, $n_e \sim 1.7 \times 10^{14}/cm^2$, respectively, consistent with previous reports [18,19]. Extrapolating this sheet density to a volume density results in an estimated donor concentration of $\sim 2 \times 10^{21}/cm^3$, which is nearly three orders of magnitude above the P-induced metal-insulator transition ($\sim 4 \times 10^{18}$ cm$^{-3}$) in Si and over five orders of magnitude greater than the background doping in the surrounding Si [23].

To complete the device, metal contacts to the donor nanostructures must be fabricated. Here, and in previous works, this is done by a two-step *ex situ* electron beam lithography (EBL) process [5,6]. The first step produces etch holes through the epitaxial layer to the donor layer. The second EBL step yields metal wires and bond pads via a liftoff step.

The key challenge to completing a device is that accurate placement of the etch holes and metal requires knowledge of the location of the buried donor structure with respect to suitable EBL registration markers. Previous works have achieved registration with accuracy on the order of a few hundred nanometers via two different methods, step-flow engineered Si etch marks or a scanning electron microscope to view the STM tip, described in detail elsewhere [5,6]. These techniques pose some challenges that make other new potential routes to registration desirable.

Here, we demonstrate a new approach to register the buried donor nanostructures. First, we show that SCM can detect and image the buried donor nanostructures produced by the STM process. Then we use the SCM to register the donor structures' location with respect to suitable EBL alignment marks in order to accurately place etch holes and metal bond pads with respect to the STM-defined structures.

The SCM measures variations in the differential capacitance ($\partial C/\partial V$) of the MOS structure formed by a conductive probe tip, a (native) oxide, and the sample as shown in Fig. 2. Variations in local doping or dielectric properties near the sample surface modify the MOS C-V curve and $\partial C/\partial V$. The $\partial C/\partial V$ signal is measured at a fixed frequency in the range 80-100 kHz using a lock-in amplifier and an AC voltage amplitude of 1-2 V. The SCM data consists of a $\partial C/\partial V$ amplitude and phase (sensed by the lock-in amplifier). The amplitude probes the local slope of the C-V curve, while the phase reveals the predominant doping type. SCM is known to be sensitive to variations of $<10^{15}/cm^3$ in bulk dopant concentration in Si with 10-nm-scale spatial resolution [20,21]. We do SCM using a Veeco-DI AFM 3000 equipped with the manufacturer's capacitance sensor and Ti-Pt coated AFM tips.

We find that SCM can image the donor nanostructure with strong contrast (S:N ~ 10:1, BW=30 Hz), Fig. 3. The SCM images of the buried donor device are shown along with the contact mode atomic force microscopy (AFM) images acquired simultaneously. The shape of the donor nanostructure, i.e. the room temperature electron distribution, is consistent with the shape of the H-lithographic pattern used to form the device, Fig. 1 (j). In addition, the SCM amplitude signal in the donor-doped regions is not DC bias-dependent (±2V) and much smaller than on the surrounding lightly-doped substrate, consistent with the expectation that the donor region is so highly doped that its SCM response is qualitatively similar to metal (for which $\partial C/\partial V = 0$). It is not possible to calibrate $\partial C/\partial V$ in Farads/Volt, so data is labeled a.u. The SCM phase undergoes a shift in sign over the donor-doped regions, consistent with a change from p-type to n-type doping going from the substrate to the donor doped structure.

In order to utilize the SCM for registration of the buried donor structure, we add standard Ti-Au EBL alignment crosses in the general vicinity (15 ± 5 µm) of the donor nanostructure, Fig. 4 (a). The location for the Ti-Au cross is chosen from the optical microscope image, Fig. 2 (g,h), and need only be accurate to a few microns. The Ti-Au crosses provide high contrast in the SCM, Fig. 4 (b,c) and sharp edges for the EBL process step. By acquiring an SCM/AFM image of the cross and nearby donor nanostructure, we locate the donor nanostructure with respect to the cross with ~300 nm accuracy, Fig. 4 (c). The coordinates guide subsequent EBL steps shown in the process. Subsequent EBL lithography steps have imprecision <20 nm with respect to the Ti-Au cross.

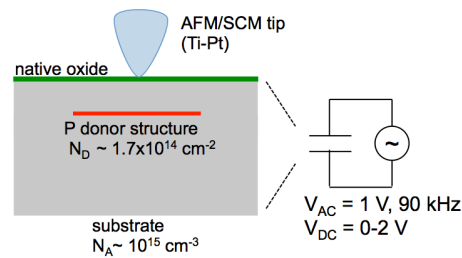

**Figure 2 Schematic for SCM measurement of buried donor structure.**

Metal contacts to the delta-doped layer are formed by first defining 100 nm diameter circular holes in a 240 nm thick ZEP mask using EBL and then dry etching the silicon to a timed target depth of 100 nm, Fig. 4 (d). SCM imaging after the etch confirms accurate placement of the holes over the buried donor nanostructure, Fig. 4 (e,f). The etched hole pattern is clearly visible along with the buried donor nanostructure, Fig. 4 (f). The etch is followed by a subsequent EBL patterning for the metal contact pads and electron-beam deposition of 150 nm of Al, Fig. 4 (g, h). The accurate placement of the metal layer is confirmed by scanning electron microscopy, Fig. 4 (i), where the metal-filled etch holes are visible in the metal layer. An overlay of the SEM and SCM images illustrates the position of all three (donor, holes, metal) device layers, Fig. 4 (j).

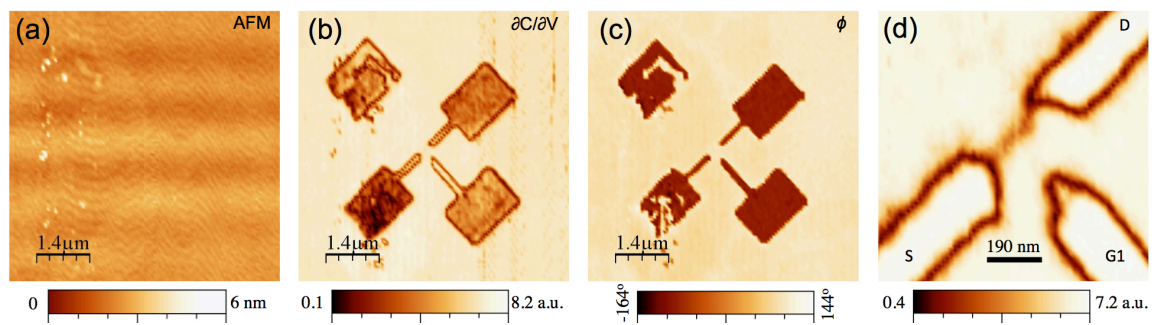

**Figure 3 Simultaneous AFM and SCM images of the SET buried donor structure. (a) Contact mode AFM image over the donor region showing the flat topography (1 nm roughness). (b) SCM amplitude ($\partial C/\partial V$) over the donor structure. (c) SCM phase signal (d) SCM image of the active region of the donor device. Although the SCM resolution is insufficient to distinguish the exact structure of the active region, the shape is consistent with donor doping resembling the lithographic pattern in Fig. 2 (i,j).**

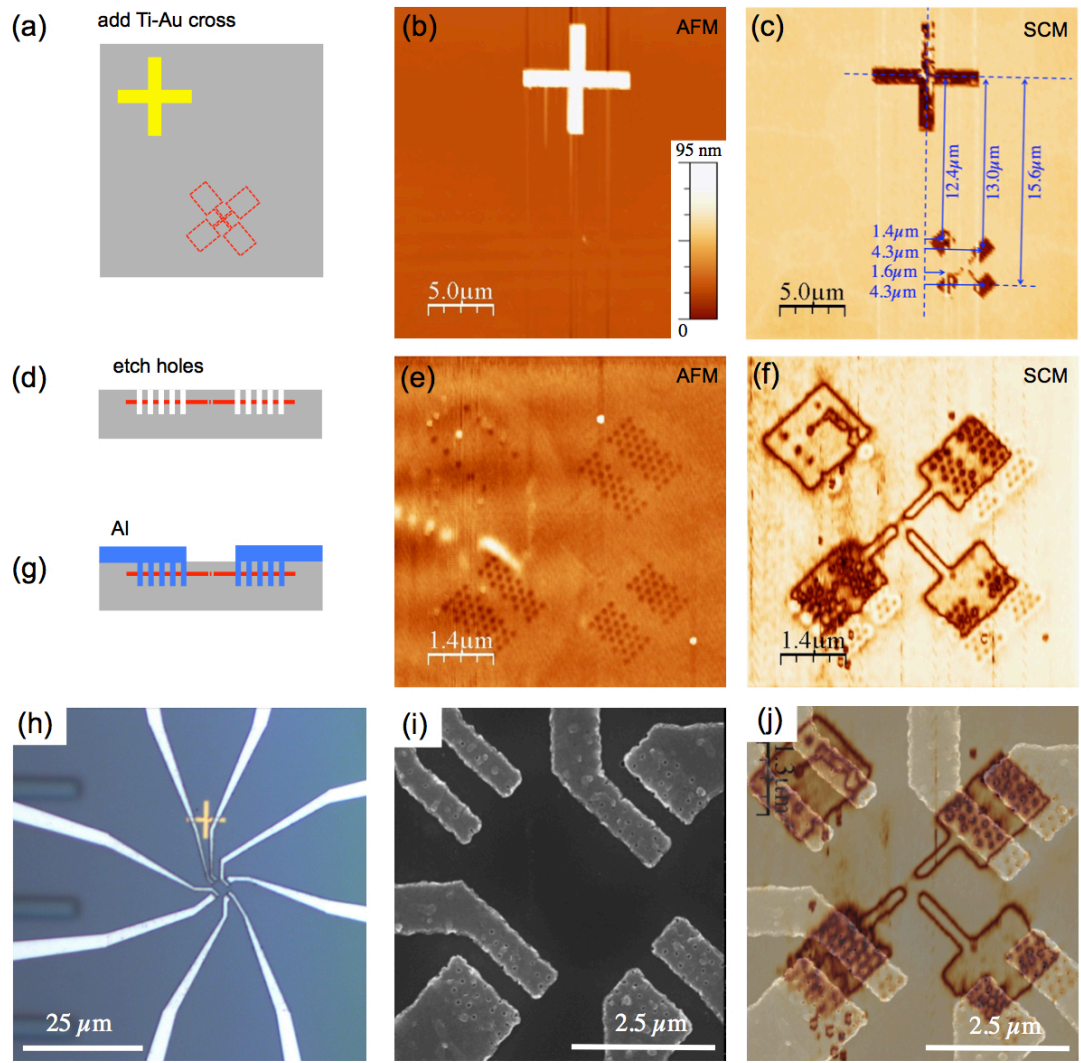

**Figure 4** Process flow for contacting buried donor structure. (a) Placement of a Ti-Au EBL alignment mark. (b) AFM of Ti-Au cross and (c) simultaneous SCM allowing registration, indicated in blue, of the buried donor structure. (d) Formation of etch holes and (e) AFM and (f) SCM of the etch holes revealing their location with respect to the donor structure. (g) Addition of Al metal wires and (h) an optical microscope image of the completed device. (i) SEM image of the metal. (j) Composite SEM and SCM image showing metal relative to the donor structures.

Another confirmation of accurate placement of the contacts to the donor nanostructure is indicated by low temperature (T=4 K) transport measurements. Since electrical carriers in the low-doped Si surrounding the nanostructures are frozen out at 4 K [23], a finite resistance between the contacts indicates successful placement of the contacts to the donor layer. Each donor contact pad has been connected to two wires labeled *a* and *b*. Fig. 5 shows I-V measurements of conduction between wires across each donor contact pad (from wire a to wire b). The contacts are ohmic with series resistances on the order of 10-200 k$\Omega$, comparable to previous reports using the STM device fabrication approach [6].

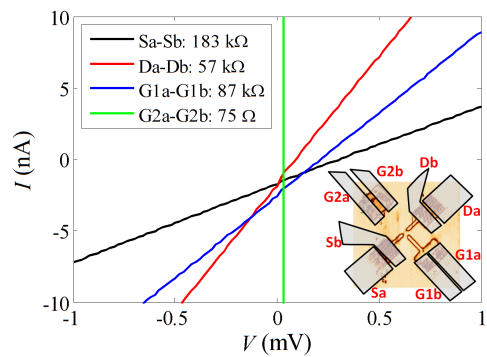

**Figure 5 Transport measurements at 4 K showing the current measured due to an applied voltage across the bond pads. Both the linear trend and the magnitude of the current are consistent with Ohmic contact to the donor delta doped layer. Inset shows the layout of the device and the labels of the bond pads for the corresponding I-V curves indicated in the legend. The 75 Ω resistance for G2a-G2b is due to an electrical short caused by the Ti-Au alignment cross (see Fig. 4 (h)).**

**Discussion**

To directly check the accuracy of our registration technique, we form composite SCM+SEM images, Fig. 4 (j), that show the donor, etch hole, and metal layers simultaneously. The deviation of the metal and etch holes from their target location above the donor layer is about 300±50 nm as determined from Fig. 4 (f) and (j), where the center line of the rectangular arrays of etch holes are clearly shifted toward the lower right hand corner of the image with respect to the underlying rectangular doped contact pads. Several devices have consistently suffered from similar placement errors. Calibration, hysteresis, and drift effects of the SCM scanner all contribute to registration errors that lead to misplacements of the etch holes and metal. By controlling for such misplacements in future devices, we expect to obtain placement errors <100 nm. The ultimate resolution of the SCM is believed to be presently limited to ~10 nm at room temperature [18], and the accuracy of our EBL tool is approximately ~20 nm, which establish an approximate ultimate lower bound on the scale of placement errors that we expect to be achievable with our registration technique.

**Summary**

SCM can image atomic-precision buried donor nanostructures fabricated by a STM-based technique recently used to demonstrate single-donor transistors [1-10]. The SCM provides nondestructive characterization of the donor nanostructures, e.g. doping level and shape of the dopant distribution (room temperature electron distribution). We have shown how to use the SCM images to accurately align subsequent device layers, using EBL steps, to the donor nanostructures. We show alignment with 300 nm accuracy. It is likely that with additional improvements in calibration of the SCM scanner, it will be possible to place metal layers on the surface with sub-100 nm precision relative to buried donor devices. Such precision is roughly an order-of-magnitude smaller than that achieved with other existing methods and will facilitate much more precise placement of surface electrodes for functions such as enhancement gates to form field effect transistors or electrodes to manipulate the electrostatic or electrodynamic (e.g., electron or nuclear spin resonance) environment of a donor based device.

**Acknowledgement**

The authors would like to thank Michelle Y. Simmons (UNSW), John Randall, Josh Ballard and James Owen (Zyvex Labs) for helpful discussions. This work was performed, in part, at the Center for Integrated Nanotechnologies, a U.S. DOE, Office of Basic Energy Sciences user facility. The work was supported by Sandia's Laboratory Directed Research and Development Program. Sandia National Laboratories is a multi-program laboratory operated by Sandia Corporation, a Lockheed-Martin Company, for the U. S. Department of Energy under Contract No. DE-AC04-94AL85000.